\documentstyle[12pt,epsfig]{article}
\renewcommand{\baselinestretch}{1.22}
\begin{document}
\newcommand{\vecvar}[1]{\mbox{\boldmath$#1$}}
%
%hep-ph/0012331

\begin{flushright}
  December, 2000 \ \ \\
  OU-HEP-374 \ \\
%  hep-ph/*******
\end{flushright}
\vspace{0mm}
\begin{center}
\large{Chiral-odd distribution functions
in the chiral quark soliton model}
\end{center}
\vspace{0mm}
\begin{center}
M.~Wakamatsu\footnote{Email \ : \ wakamatu@miho.rcnp.osaka-u.ac.jp}
\end{center}
\vspace{-4mm}
\begin{center}
Department of Physics, Faculty of Science, \\
Osaka University, \\
Toyonaka, Osaka 560, JAPAN
\end{center}

\vspace{8mm}
%\begin{flushleft}
%\end{flushleft}

\vspace{10mm}
\begin{center}
\small{{\bf Abstract}}
\end{center}
\vspace{-1mm}
\begin{center}
\begin{minipage}{15.5cm}
\renewcommand{\baselinestretch}{1.0}
\small
\ \ \ The recent measurements of azimuthal single spin asymmetries
by the HERMES collaboration has opened up new possibility to measure
chiral-odd distribution functions through semi-inclusive 
deep-inelastic scatterings. Here, predictions are given for the
twist-2 and twist-3 chiral-odd distribution
functions of the nucleon within the framework of the chiral quark
soliton model, with full inclusion of the vacuum polarization
effects as well as the subleading $1/N_c$ corrections.
The importance of the vacuum polarization effects is 
demonstrated by showing that the so-called Soffer inequality 
holds not only for the quark distributions but also for the 
antiquark ones.            
\end{minipage}
\end{center}

\normalsize
\vspace{5mm}

It is widely believed that the transversity distribution function 
$h_1(x)$ provides us with valuable information for our thorough 
understanding of internal nucleon spin structure \cite{JJ92}.
Because of its chiral-odd nature, however, it does not appear 
in the standard deep-inelastic inclusive cross sections
(at least in the chiral limit). The only practical way to determine
it has been believed to use Drell-Yang processes.
Recently, the HERMES collaboration proved for the first
time that it is practically feasible to measure chiral-odd
distribution functions by making use of the so-called Collins
mechanism \cite{Coll93} in the semi-inclusive electro-pion
productions \cite{HERMES00}.
The first HERMES measurement of the azimuthal single target-spin
asymmetries for charged pions has been done by using longitudinally
polarized proton target. From the theoretical viewpoint, much simpler 
would be the measurement of the similar asymmetry for the target
polarized transversely to the incident electron beam, since it 
enables us to determine the transversity distribution $h_1 (x)$
more directly and efficiently.
Such measurement will in fact be performed soon.

  In view of this very exciting experimental situation, it is highly 
desirable to  give some reliable predictions for the chiral-odd
distribution functions of the nucleon. Naturally, complete
understanding of the parton distribution functions (PDF) need to
solve nonperturbative QCD dynamics.
Unfortunately, the direct evaluation of the PDF
in the lattice QCD simulation is not possible yet.
At the present moment, we must therefore rely upon some effective
theory of QCD. We would claim that, among many of them,
most promising would be the chiral quark soliton model
(CQSM) \cite{DPP88}-\cite{WK99}.
In fact, a prominent feature of the CQSM is that it can
simultaneously explain two biggest findings in the recent
experimental studies of high-energy deep-inelastic scattering
observables, i.e. the EMC measurement in 1988 \cite{EMC88}, which 
dictates unexpectedly small quark spin fraction of the nucleon, 
and the NMC measurement in 1991, which has revealed the excess of 
$\bar{d}$ sea over the $\bar{u}$ sea in the proton \cite{NMC91}.
Furthermore, its unique prediction, i.e. the flavor asymmetry of
the spin-dependent sea-quark distribution $\Delta \bar{u}(x) - 
\Delta \bar{d}(x) > 0$ \cite{WW00A},\cite{DGPW00},
seems to win some semi-phenomenological and/or semi-theoretical
support \cite{ASYM}, although not entirely definite yet.

What is the chiral quark soliton model, then?
For large enough $N_c$, a nucleon in this model is thought to be a
composite of $N_c$ valence quarks and infinitely many Dirac sea
quarks bound by the self-consistent pion field of hedgehog
shape \cite{DPP88}.
After canonically quantizing the spontaneous rotational motion of
the symmetry breaking mean field configuration, we can perform
nonperturbative evaluation of any nucleon observables with full
inclusion of valence and deformed Dirac-sea quarks \cite{WY91}.
This incomparable feature of the model enables us to make
a reasonable estimation not only of quark distributions but also of
antiquark ones, as we shall demonstrate later.
Finally, but most importantly, only 1 parameter of the model
(dynamically generated quark mass $M$) was already fixed by low energy
phenomenology, which means that we can give {\it parameter-free
predictions} for parton distribution functions
at the low renormalization scale.
There already exist some investigations of the chiral-odd distribution
functions based on the CQSM (or the Nambu-Jona-Lasinio soliton model).
The first investigation of the transversity distribution $h_1 (x)$ in 
the CQSM was done by Pobylitsa and Polyakov \cite{PP96}.
Although pioneering, their investigation should be taken as qualitative
in several respects.
First, the effect of vacuum polarization is entirely neglected.
Second, only the isovector combination of $h_1 (x)$, i.e.
$h_1^u (x)-h_1^d (x)$, was estimated, and the isocalar piece
$h_1^u (x) + h_1^d (x)$, which arises only at the subleading order 
of $1/N_c$ expansion, was left over.
On the other hand, Gamberg et al. investigated both of
$h_1^{u/d}(x)$ and $h_L^{u/d}(x)$, but also within the
``valence-quark-only'' approximation \cite{GRW98}. 
We shall later show that this approximation does
not make full use of the potential power of the model and cannot
be justified.
The purpose of the present investigation is to give more complete
predictions for the chiral-odd distribution functions with full
inclusion of the vacuum polarization effects as well as the subleading 
$1/N_c$ corrections.

We start with the standard definition of the
chiral-odd spin-dependent PDF \cite{JJ92} :
\begin{eqnarray}
 &\,& \!\!\! h_1^{(I=0 / I=1)} (x) \ \equiv \
 h_1^u (x) \pm h_1^d (x) \ = \ \frac{1}{4 \,\pi} \,
 \int_{- \infty}^{\infty} \,dz_0 \,\, e^{i x M_N z_0} \nonumber \\
 \ \ \ \ \ &\,& \times \ \langle N (\vecvar{P}=0) S_\perp \,\vert \,
 \psi^\dagger (0)
 \left( 1 + \gamma^0 \gamma^3 \right) \,\gamma_\perp \,\gamma_5 \,
 \left\{ \begin{array}{c} 1 \\ \tau_3 \\ \end{array} \right\}
 \,\psi(z) \,\vert \,N(\vecvar{P}=0) S_\perp \rangle
 \vert_{z_3 = - z_0, z_\perp = 0} , \hspace{10mm} \\
 &\,& \!\!\! h_L^{(I=0 / I=1)} (x) \ \equiv \
 h_L^u (x) \pm h_L^d (x) \ = \ \frac{1}{4 \,\pi} \,
 \int_{- \infty}^{\infty} \,dz_0 \,\, e^{i x M_N z_0} \nonumber \\
 \ \ \ \ \ &\,& \times \ \langle N (\vecvar{P}=0) S_z \,\vert \,
 \psi^\dagger (0) \,
 \gamma^3 \,\gamma_5 \,
 \left\{ \begin{array}{c} 1 \\ \tau_3 \\ \end{array} \right\}
 \,\psi(z) \,\vert \,N(\vecvar{P}=0) S_z \rangle
 \vert_{z_3 = - z_0, z_\perp = 0} .
\end{eqnarray}
What we must evaluate here is the nucleon matrix elements of quark
bilinear operators with light-cone separation. As fully discussed
in \cite{DPPPW96},\cite{WK99}, on the basis of the path integral
formulation of the CQSM, such nonlocality effects in time and
spatial coordinates can be
treated in a satisfactory way. Omitting the detail, we just recall
here the fact that, within the framework of the CQSM,
the isoscalar and isovector distributions have
totally dissimilar theoretical structure since they
have different dependence on the collective angular velocity $\Omega$ :
\begin{eqnarray}
 h_{1/L}^u (x) \ + \ h_{1/L}^d (x) &{\sim}& \,N_c \,\,
 O (\Omega^1) \hspace{24mm} \sim \ O (N_c^0) \, ,\\
 h_{1/L}^u (x) \ - \ h_{1/L}^d (x) &{\sim}& N_c \,
 [ \,O(N_c^1) \ + \ O(\Omega^1) \,] \ \sim \
 O(N_c^1) \ + \ O(N_c^0) ,
\end{eqnarray} 
where use has been made of the fact that $\Omega$ scales as $1/N_c$.
A noteworthy feature here is the existence of the subleading
$1/N_c$ correction to the isovector combination of the distributions.
Its importance is already known from a similar analysis for the
isovector longitudinally polarized distribution \cite{WK99} or its
first moment, i.e. the isovector axial coupling constant of the
nucleon \cite{WW93}.

We summarize in Fig.1 the theoretical predictions for the chiral-odd
distribution functions $h_1 (x)$ and $h_L (x)$. Fig.1(a) and Fig.1(b)
respectively stand for the isoscalar and isovector combinations of 
$h_1^u (x)$ and $h_1^d (x)$, while (c) and (d) represent the similar
combinations of $h_L^u (x)$ and $h_L^d (x)$.
Here, the distribution functions with negative $x$ should be
interpreted as aitiquark ones according to the rule :
\begin{equation}
 h_{1/L}^u (-x) \ \pm \ h_{1/L}^d (-x) \ = \ - \,\,
 [ \,h_{1/L}^{\bar{u}}(x) \ \pm \ 
 h_{1/L}^{\bar{d}}(x) \,] \hspace{10mm} (0<x<1) .
\end{equation}
In all the figures, the long-dashed curves peaked around 
$x \sim 1/3$ represent the contributions of $N_c$ valence quarks,
while the dash-dotted curves are those of Dirac-sea quarks in the
hedgehog mean field.
The sums of these two contributions are denoted by solid curves.
Concerning the transversity distributions $h_1 (x)$, one sees that the
effects of Dirac-sea quarks are not very large.
One also notices that both of $h_1^u (x) + h_1^d (x)$ and 
$h_1^u (x) - h_1 ^d (x)$ have fairly small support in the negative
$x$ region, which means that the transversity distributions do not
have significant antiquark components, in contrast to the unpolarized
as well as the longitudinally polarized distributions.
Turning to the twist-3 distribution $h_L (x)$, one finds that the
effect of Dirac-sea quarks are sizably large.
Nevertheless, an interesting feature is that a considerable
cancellations occurs between the contributions of valence quarks
and of Dirac-sea quarks in the negative $x$ region.
As a consequence, the antiquark components are not so significant also
in the case of $h_L (x)$.

\begin{small}
\begin{figure}[htbp] % fig 1
\centerline{\epsfig{file=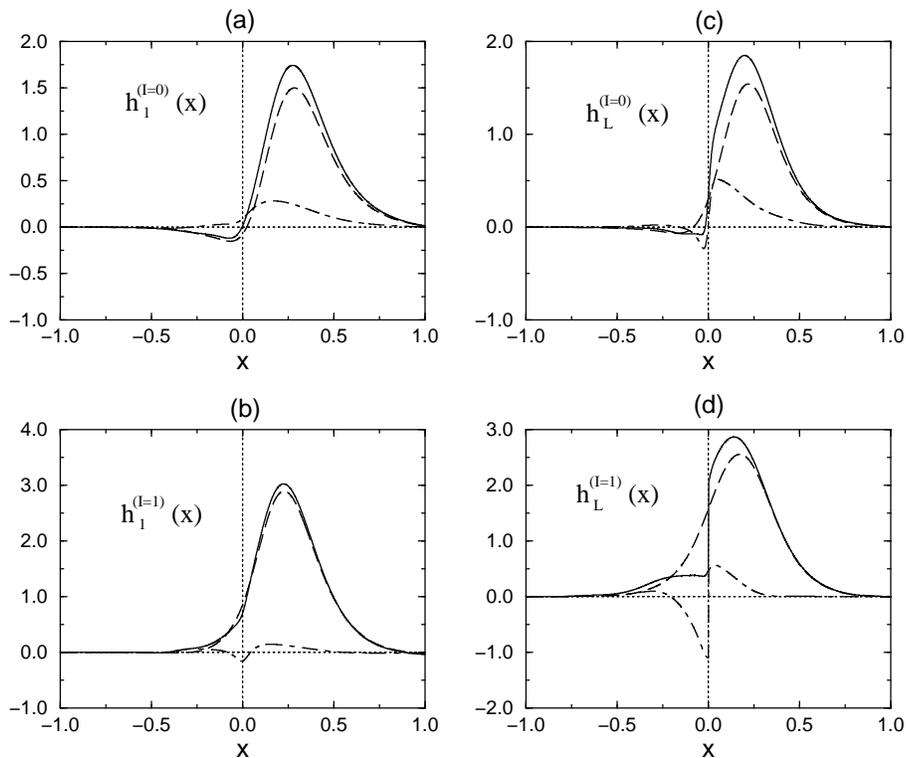,width=12.0cm}}
%\psbox[width=12.0cm]{transpdf.eps}
%\renewcommand{\baselinestretch}{1.00}
\caption{Theoretical predictions for the chiral-odd distributions
  $h_1 (x)$ and $h_L (x)$. (a) and (b) respectively represent the
  isoscalar and isovector parts of $h_1 (x)$, while (c) and (d)
  are the isoscalar and isovector parts of $h_L (x)$. In all
  the figures, the long-dashed and dash-dotted curves
  denote the contributions of $N_c$ valence quarks and of the
  Dirac-sea quarks, whereas the solid curves are the sum of these
  two contributions.}
\end{figure}
\end{small}

The transversity distribution $h_1^q (x)$ is of special importance,
since it forms the set of twist-2 distribution functions, together with
the unpolarized distributions $f_1^q (x)$ and the longitudinally 
polarized ones $g_1^q (x)$.
It is known that these distribution functions must satisfy the
so-called Soffer inequality \cite{Soffer95} expressed in the following
form :
\begin{equation}
 \vert h_1^q (x) \vert \ \leq \ \frac{1}{2} \,\left( \,
 \pm \,f_1^q (x) \ + \ g_1^q (x)\right) \hspace{10mm}
 \left( x > 0, \,x < 0 \right) .
\end{equation}

Here, the plus sign of $f_1^q (x)$ corresponds to the region $x > 0$,
while the minus sign to $x < 0$.
(Note that the inequality with negative argument gives a relation
for the antiquark distributions.)
Now the question is whether the predictions of the CQSM fulfill
this general inequality or not. Fig.2 shows that, if one includes the
vacuum polarization effects properly, the Soffer inequality is
well satisfied for both of the $u$-quark and the $d$-quark.
On the other hand, if one neglects the Dirac-sea contributions, it is
badly broken for the antiquark distributions.
An important lesson learned from this observation is that the field
theoretical nature of the model, i.e. the proper inclusion of the
{\it vacuum polarization effects}, plays essential roles in giving
reasonable predictions for {\it antiquark distributions}. Another lesson is
that the frequently-used saturation Ansatz of this inequality
for estimating $h_1^q (x)$ is not justified. This seems reasonable to
us, since the magnitude of $f_1^q (x)$ is much larger than $g_1^q (x)$
and since $h_1^q (x)$ is closer to $g_1^q (x)$ rather than $f_1^q (x)$.

\begin{small}
\begin{figure}[htbp] % fig 2
\centerline{\epsfig{file=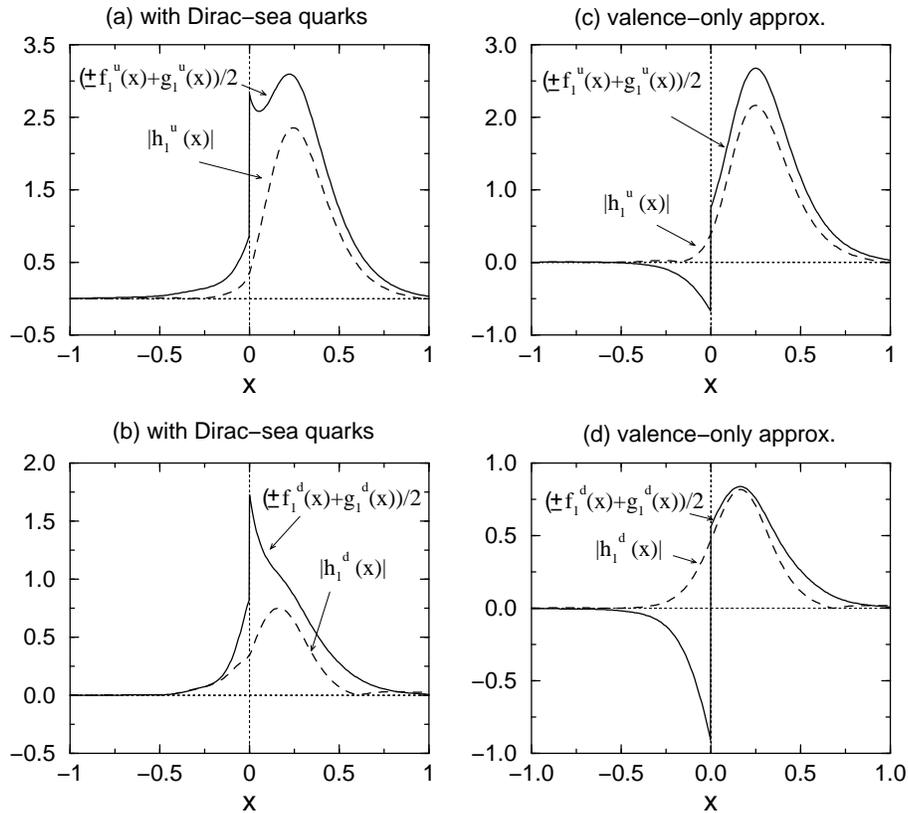,width=12.0cm}}
%\psbox[width=12.0cm]{sineq.eps}
%\renewcommand{\baselinestretch}{1.00}
\caption{Theoretical check of the Soffer inequality. (a) and (b)
  are the results obtained with full inclusion of the vacuum
  polarization effects, while (c) and (d) are obtained with
  ``valence-quark-only'' approximation. In all the figures, the
  solid and dashed curves represent the r.h.s and the l.h.s. of the
  inequality (6), respectively.}
\end{figure}
\end{small}

As already touched upon, HERMES group recently carried out a very
interesting measurements on single target-spin asymmetries in
semi-inclusive pion electroproduction \cite{HERMES00}.
What they have measured are
the azimuthal asymmetries, i.e. the asymmetries of the cross sections
depending on the azimuthal angle $\phi$ with respect to the lepton
scattering plane :
\begin{equation}
  A_{UL}^W \ = \ \frac{\int d \phi \,dy \,\,W(\phi) \,\,(
  d \sigma^{+} / S_H^{+} \,\,dx \,dy \,d \phi \ - \
  d \sigma^{-} / S_H^{-} \,\,dx \,dy \,d \phi)}{ \frac{1}{2} \,\,
  \int d \phi \,dy \,\,(
  d \sigma^{+} / S_H^{+} \,\,dx \,dy \,d \phi \ + \
  d \sigma^{-} / S_H^{-} \,\,dx \,dy \,d \phi)} ,
\end{equation}
with $W(\phi) = \sin \phi$ or $\sin 2 \phi$ and with $S_H^{\pm}$
being the nucleon polarization.
The theoretical analyses of these azimuthal asymmetries were already
performed by several authors \cite{Kotz95}-\cite{OBSN00}.
Here we use the expression given in \cite{OBSN00} :
\begin{eqnarray}
  A^{\sin \phi} &\simeq& \,\frac{2 \,M_h}{\langle \,P_{h T} \rangle}
  \cdot \langle \frac{| P_{h T} |}{M_h} \,\,\sin \phi \,\rangle
  \ \ \ \ \ \, =  \ \frac{2 \,M_h}{\langle P_{h T} \rangle} \cdot
  \frac{I_2 (x,y,z)}{I_1 (x,y,z)} , \\
  A^{\sin 2 \,\phi} &\simeq&
  \frac{2 \,M \,M_h}{\langle \,P_{h T}^2 \rangle}
  \cdot \langle \frac{{| P_{h T} |}^2}{M \,M_h} \,\,
  \sin 2 \,\phi \rangle
  \ =  \ \frac{2 \,M \,M_h}{\langle P_{h T}^2 \rangle} \cdot
  \frac{I_3 (x,y,z)}{I_1 (x,y,z)} ,
\end{eqnarray}
where
\begin{eqnarray}
  I_1 (x,y,z) &=& \frac{1}{2} \,\,\left[ 1 + {(1-y)}^2 \right] \,\,
  \sum_a \,e_a^2 \,\,x \,f_1^a (x) \,\,D_1^a (z) , \\
  I_2 (x,y,z) &=& 2 \,(2-y) \,\frac{M}{Q} \,\,\sum_a \,e_a^2 \,\,
  \left\{ \,x^2 \,h_L^a (x) \cdot z \,
  H_1^{\perp (1)a} (z) \ - \
  x \,h_{1L}^{\perp (1)a} (x) \cdot 
  \tilde{H}^a (z) \,\right\} \nonumber \\
  &+& \!\! \sqrt{1-y} \,\,(1-y) \,\,\frac{2 \,M \,x}{Q} \,\,
  \sum_a \,e_a^2 \,\,x \,h_1^a (x) \cdot z \,H_1^{\perp (1)a} (z), \\
  I_3 (x,y,z) &=& 4 \,(1-y) \,\,\sum_a \,\,e_a^2 \,
  \,x \,h_{1L}^{\perp (1)a} (x) \cdot z^2 \,H_1^{\perp (1)a} (z),
\end{eqnarray}
with $x = Q^2 / 2 (P \cdot q), y = (P \cdot q) / (P \cdot k_1)$ and
$z = (P \cdot P_h) / (P \cdot q)$ in the notation of \cite{OBSN00}.
One realizes that the asymmetries depend on 4 distribution functions
$f_1 (x), h_1 (x), h_L (x), h_{1L}^{\perp(1)} (x)$ and 3 fragmentation
functions $D_1 (z), H_1^{\perp(1)} (z), \tilde{H} (z)$.
Here $D_1 (z)$ is the familiar spin-independent fragmentation
function, while $H_1^{\perp(1)} (z)$ is $(k_T^2 / 2 M_h^2 )$-weighted
fragmentation function defined by
\begin{equation}
 H_1^{\perp(1)a} (z) \ \equiv \ \int \,d^2 \vecvar{k}^\prime_T \,\,
 \frac{| \vecvar{k}_T |^2}{2 \,M_h^2} \,\,
 H_1^{\perp a} (z,k_T^\prime) ,
\end{equation}
with $\vecvar{k}_T^\prime = - z \vecvar{k}_T$, where
$H_1^{\perp q} (z,\vecvar{k}_T^\prime)$ is a $T$(time-reversal)-odd
leading twist fragmentation function, giving the probability of a
spinless or unpolarized hadron to be created from a transversely
polarized scattered quark. The remaining
subleading $T$-odd fragmentation function $\tilde{H}^a (z)$ is
known to be constrained by the relation $\tilde{H}^a (z) =
z \,\frac{d}{dz} \left( z H_1^{\perp(1)a} (z) \right)$ due to the
Lorentz invariance \cite{TM95}.
To estimate these T-odd fragmentation functions,
we follow \cite{OABK98}-\cite{OBSN00} and use the Collins Ansatz
\cite{Coll93}
\begin{equation}
 A_C (z,\vecvar{k}_T) \ \equiv \ \frac{|\vecvar{k}_T|}{M_h} \cdot
 \frac{H_1^{\perp} (z,\vecvar{k}_T^2)}{D_1 (z,\vecvar{k}_T^2)}
 \ = \ \eta \,\,\frac{M_C |\vecvar{k}_T|}{M_C^2 + \vecvar{k}_T^2},
\end{equation}
with the parameters $M_C = 2 \,m_\pi$ and $\eta = 1.0$.

Coming back to the distribution functions, we need 4 functions
$f_1 (x), h_1 (x), h_L (x)$ and $h_{1L}^{\perp(1)} (x)$.
In the following, simply by using the GRV parameterization of the
unpolarized distribution $f_1^a (x)$ \cite{GRV95},
we concentrate on the chiral-odd
distributions. Here, $h_1^a (x)$ is the familiar transversity
distribution function. The function $h_L (x)$ consists of the twist-2
part, which is expressed by $h_1^a (x)$, and the interaction dependent
part or the genuine twist-3 part, as
\begin{equation}
  h_L^a (x) \ = \ 2 \,x \,\,\int_x^1 \,\,dy \,\,
  \frac{h_1^a (y)}{y^2} \ + \ 
  \bar{h}_L^a (x) .
\end{equation} 
On the other hand, the third function $h_{1L}^{\perp(1)a} (x)$
is expressed with use of $h_1^a (x)$ and $h_L^a (x)$ as
\begin{equation}
  h_{1L}^{\perp (1)a} (x) \ = \ \int_x^1 \,\,dy \,\,\left[ \,
  h_L^a (y) \ - \ h_1^a (y) \,\right] .
\end{equation}
In the absence of powerful theories, the following two approximations
have been frequently used for estimating these functions. The first is
the twist-2 or the Wandzura-Wilczek type approximation for $h_L^a (x)$,
i.e. the approximation to set $\bar{h}_L^a (x) \simeq 0$.
The second approximation is to assume approximate equality of
$h_1^q (x)$ and $h_L^a (x)$, which leads to $h_{1L}^{\perp(1)a} (x)
\simeq 0$. This second approximation is advocated by HERMES group,
by the reason that it seems consistent with small $\sin 2 \phi$
asymmetry obtained by their experiment \cite{OABK98}-\cite{OBSN00}.

Since the CQSM can give some reasonable predictions for both of
$h_1 (x)$ and $h_L (x)$, the other distributions can also be
determined by using (16). Fig.3 summarizes the theoretical
predictions of the CQSM for these chiral-odd distributions with
different flavors. The distributions shown here corresponds to the
energy scale of $Q^2 = 2.5 \,\mbox{GeV}^2$, an average value for
the HERMES kinematical region. The scale dependencies of the
distribution functions are taken into account in the following way.
First, to include the scale dependencies of $h_1 (x)$ and the
twist-2 part of $h_L (x)$, we use the Fortran program of
leading-order DGLAP equation \cite{HKM98}. On the other hand, the scale
dependence of the twist-3 part of $h_L (x)$, i.e. $\bar{h}_L (x)$,
is taken into account by solving the leading-order DGLAP type  
equation obtained in the large $N_c$ limit \cite{BBBKT96},\cite{KK97}.
The starting energy of these evolutions is taken to be
$Q^2 = 0.23 \, \mbox{GeV}^2$ \cite{WW00B}.
From Fig.3(a) and 3(b), one reconfirms sizable differences between the
two distributions $h_1 (x)$ and $h_L (x)$, especially at lower
values of $x$. A general tendency is that $h_L (x)$ is concentrated
in the lower $x$ region as compared with $h_1 (x)$. We recall that
this characteristic is also observed in the MIT bag model
predictions for the same distributions. In fact, the chiral-odd
distributions $h_1 (x)$ and $h_L (x)$ in both models are not
extremely different as far as the dominant $u$-quark distribution
in the proton is concerned.
(This is not necessarily true for the corresponding
$d$-quark distributions as well as the $\bar{u}$ and $\bar{d}$ ones.
We also recall that the situation is quite different for the other two
twist-2 distributions, i.e. the longitudinally polarized one and
the unpolarized one.) The sizable difference between $h_1 (x)$ and
$h_L (x)$ indicates that the Ansatz $h_1 (x) \simeq h_L (x)$
adopted by the HERMES group may not be necessarily justified.
To verify it, we plot in Fig.3(d) $h_{1L}^{\perp(1)} (x)$
obtained from (16). One sees that $h_{1L}^{\perp(1)} (x)$ for
$u$-quark has in fact nonnegligible magnitude, although the
distributions for $d, \bar{u}$ and $\bar{d}$ are pretty small.
We also show in Fig.3(c) the twist-3 part of $h_L (x)$ obtained
with (15). They are generally small but certainly nonnegligible
especially in the small $x$ region.

\begin{small}
\begin{figure}[htbp] % fig 3
\centerline{\epsfig{file=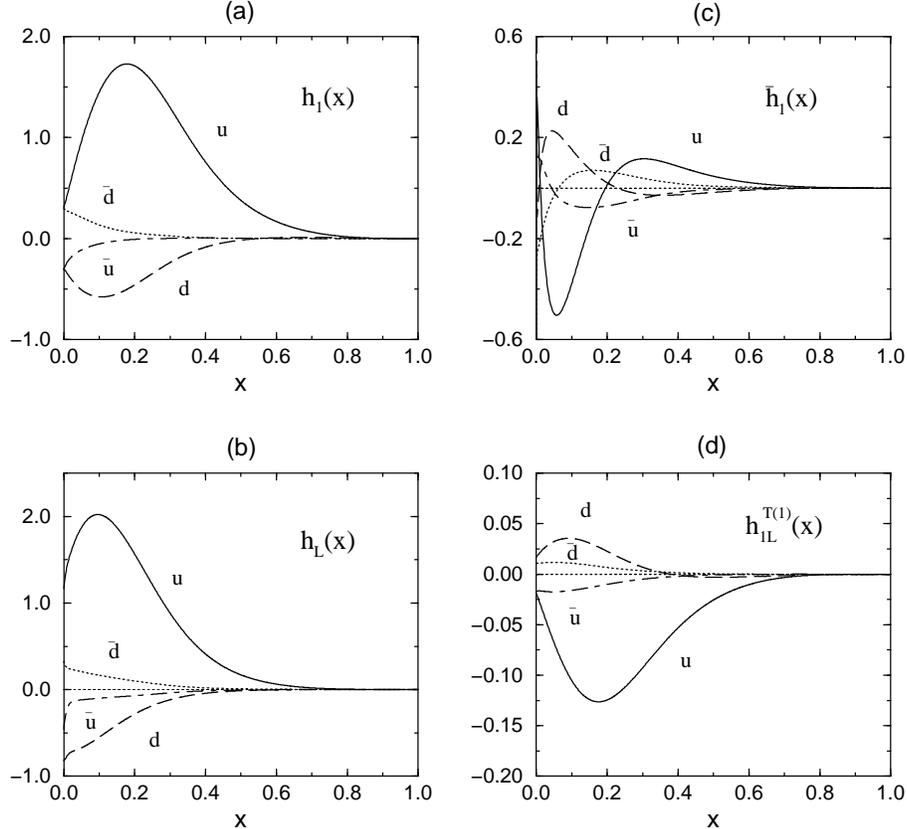,width=12.0cm}}
%\psbox[width=12.0cm]{h1LTb.eps}
%\renewcommand{\baselinestretch}{1.00}
\caption{Theoretical predictions for the chiral-odd distribution
  functions $h_1 (x), h_L (x), \bar{h}_L (x)$ and $h_{1L}^{\perp (1)}
  (x)$ for each flavor at the energy scale $Q^2 = 2.5 \,\mbox{GeV}^2$.}
\end{figure}
\end{small}

\begin{small}
\begin{figure}[htbp] % fig 4
\centerline{\epsfig{file=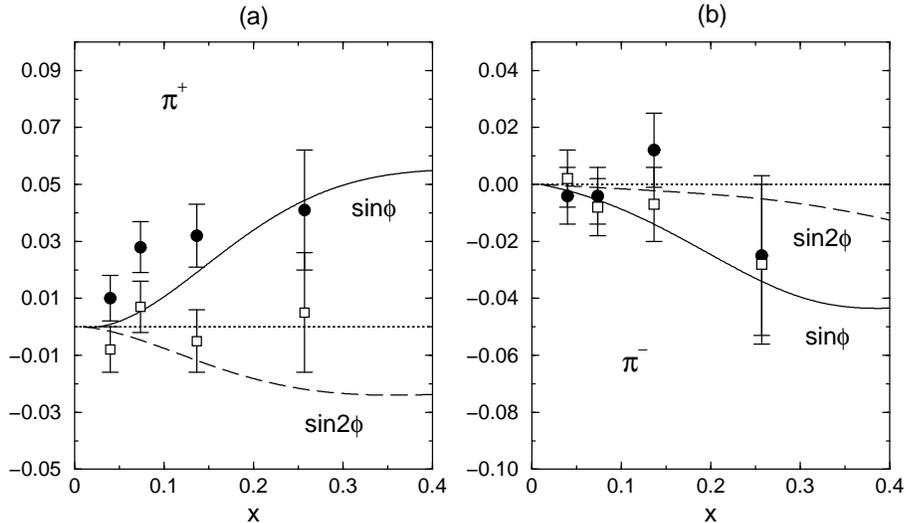,width=12.0cm}}
%\psbox[width=12.0cm]{asym.eps}
%\renewcommand{\baselinestretch}{1.00}
\caption{Theoretical predictions of the CQSM for the $\sin \phi$ and
  $\sin 2 \phi$ asymmetries in the semi-inclusive electro-pion
  productions in comparison with the HERMES experiment.}
\end{figure}
\end{small}

Fig.4 shows a preliminary comparison between the predictions of
the CQSM and the HERMES experiments for the $\sin \phi$ and
$\sin 2 \phi$ asymmetries. The theoretical curves have been
calculated as in \cite{OABK98},\cite{OBSN00} by integrating
over $y$ and $z$ in the HERMES kinematical range with the
parameter $a = 0.44 \,\mbox{GeV}$ and $b = 0.36 \,\mbox{GeV}$
by assuming Gaussian distribution for the distribution
functions and the fragmentation functions.
The main difference with the similar
analysis done in the same model \cite{EGPU00} is that we have included
all the subleading contributions into the distribution functions (no
twist-2 approximation) as well as into the fragmentation functions. 
In fact, it turns out that the subleading term containing
$\tilde{H}^a (z)$ in (11) gives sizable contributions to the
$\sin \phi$ asymmetry. One can say that the theory reproduces
the qualitative features of the data, although the uncertainties of
the present experimental data are still too large to draw any
decisive conclusion. As pointed out before, the HERMES group
advocates that the observed small $\sin 2 \phi$ asymmetry is
consistent with the Ansatz $h_1 (x) \simeq h_L (x)$, or equivalently
$h_{1L}^{\perp(1)} (x) \simeq 0$. (Note that the formula for
$\sin 2 \phi$ asymmetry is proportional to this distribution
function.) On the other hand, the CQSM as well as the MIT bag model
indicates that these two functions are rather different especially
in the small $x$ domain. Unfortunately, it seems difficult to
draw a definite conclusion only from the present HERMES data
for the $\sin 2 \phi$ asymmetry.
We certainly need more accurate experimental data.

From the theoretical viewpoint, the asymmetry obtained with a
target polarized transversely to the direction of incident electron
beam is much simpler \cite{TM95}. Such asymmetry is proportional to  
\begin{eqnarray}
 A^{\sin \phi}_{OT} &\sim&
 \frac{4 \,\pi \alpha^2 \,s}{Q^4} \,\,\vert S_T \vert \,\,(1-y)
 \,\,\sum_a \,\,e_a^2 \,\,x \,h_1^a (x) \,
 \,H_1^{\perp (1)a} (z) .
\end{eqnarray}
By measuring it, we can then get more direct information on the
transversity distribution $h_1 (x)$. Under the dominant-flavor-only
approximation for the fragmentation functions, the semi-inclusive
$\pi^+$ and $\pi^-$ productions respectively measure the
following combinations of the transversity distributions :
\begin{eqnarray}
  \pi^{+} \ &:& \ \frac{4}{9} \,\,h_1^u (x) \ + \
  \frac{1}{9} \,\,h_1^{\bar{d}} (x), \\
  \pi^{-} \ &:& \ \frac{1}{9} \,\,h_1^d (x) \ + \
  \frac{4}{9} \,\,h_1^{\bar{u}} (x).
\end{eqnarray}

\begin{small}
\begin{figure}[htbp] % fig 5
\centerline{\epsfig{file=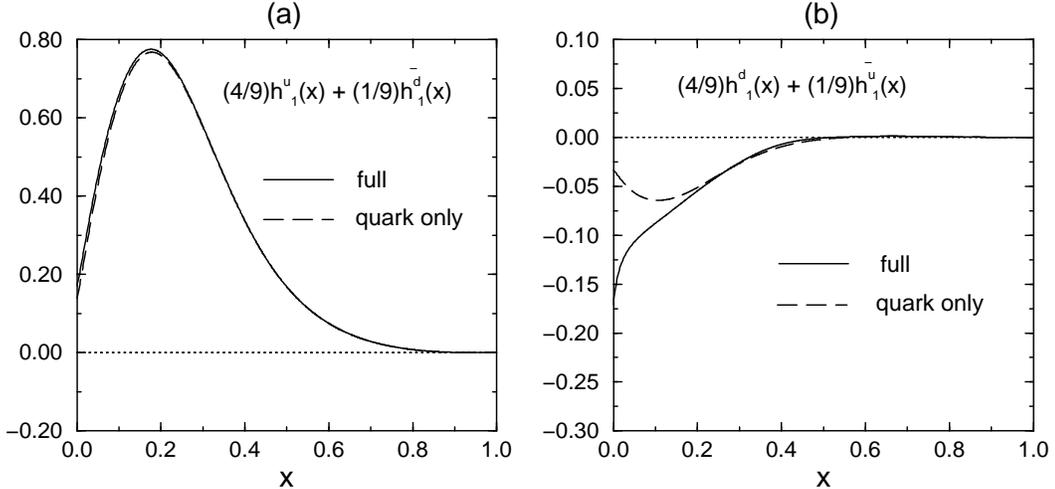,width=14.0cm,height=7.0cm}}
%\psbox[width=12.0cm]{h1pi.eps}
%\renewcommand{\baselinestretch}{1.00}
\caption{Theoretical predictions for the combinations of the
  transversity distributions,
  $\frac{4}{9} h_1^u (x) + \frac{1}{9}
  h_1^{\bar{d}} (x)$ and $\frac{1}{9} h_1^d (x) + \frac{4}{9}
  h_1^{\bar{u}} (x)$ at $Q^2 = 2.5 \,\mbox{GeV}^2$, which are to be
  measured in semi-inclusive $\pi^+$ and $\pi^-$ productions with a
  target proton polarized transversely to the direction of incident
  electron beam.}
\end{figure}
\end{small}

The solid curves in Fig. 5 stand for the full predictions of the
CQSM for these distributions at $Q^2 = 2.5 \,\mbox{GeV}^2$,
while the dashed curves here are obtained by dropping the
antiquark components in these combinations.
One sees that the first combination is quite insensitive to the
presence of the antiquark component, while the second one is
sensitive to the antiquark component at least in the small $x$
region. This indicates that $\pi^-$ semi-inclusive production may
be a useful tool to probe the role of antiquark in the transversity
distributions.

In summary, we have given theoretical predictions for the chiral-odd
distribution functions $h_1 (x)$ and $h_L (x)$ within the framework
of the CQSM, with full inclusion of the vacuum polarization effects
as well as the subleading $1 / N_c$ corrections. The importance of
the vacuum polarization effects has been demonstrated by showing
that the so-called Soffer inequality holds not only for the quark
distributions but also for the antiquark ones. The theoretical
predictions of the model for the azimuthal single spin asymmetries
in the semi-inclusive electro-pion productions are shown to be
qualitatively consistent with the corresponding HERMES data,
although we certainly need more accurate experimental data as well
as more precise knowledge about the T-odd fragmentation functions
before drawing any decisive conclusion.
We also find that the chiral-odd distribution functions have rather
different magnitudes and $x$-dependencies as compared with the
longitudinally polarized ones investigated in previous works.
Especially interesting is that the {\it transversity distribution}
have very {\it small antiquark components}. We hope that these unique
predictions of the CQSM will be tested through near future
semi-inclusive meaurements, which enables the {\it flavor} as well
as the {\it valence plus sea-quark decompositions}.

\vspace{10mm}
\noindent
\begin{large}
{\bf Acknowledgement}
\end{large}
\vspace{3mm}

The author would like to express his sincere thanks to K.A~Oganessyan
for useful discussion on the azimuthal single spin asymmetries
in the semi-inclusive electro-pion productions.

%
%  Reference
%

\renewcommand{\baselinestretch}{1.0}

\end{document}